\documentclass[groupedaddress,preprintnumbers,a4paper,twocolumn,nobalancelastpage,floatfix,aps,nofootinbib,superscriptaddress]{revtex4-1}
\usepackage{graphicx}
\usepackage{amsmath,amssymb,amsfonts}
\usepackage{hyperref}
\usepackage{bbm}
\usepackage{xspace}
\usepackage{verbatim}
\usepackage{bm}
\usepackage{color}
\usepackage{cancel}
\usepackage{ulem}
\usepackage{bbm}

\newcommand{\AddrStockholm}{
\textit{Stockholm University and The Oskar Klein Centre for Cosmoparticle Physics, Alba Nova, 10691 Stockholm, Sweden}}
\newcommand{\AddrNordita}{
\textit{Nordita, KTH Royal Institute of Technology and Stockholm University, Hannes Alfv\'{e}ns v\"{a}g 12, 10691 Stockholm, Sweden}
}
\newcommand{\AddrTexas}{
\textit{Department of Physics, The University of Texas at Austin, Austin, 78712 TX, USA}
}

\begin{document}
\title{Natural Chain Inflation}
\author{Katherine Freese} 
\email{ktfreese@utexas.edu}
\affiliation{\AddrTexas}
\affiliation{\AddrStockholm}
\affiliation{\AddrNordita}
\author{Aliki Litsa} 
\email{aliki.litsa@fysik.su.se}
\affiliation{\AddrStockholm}
\author{Martin Wolfgang Winkler}
\email{martin.winkler@su.se}
\affiliation{\AddrTexas}
\affiliation{\AddrStockholm}
\preprint{UTTG-18-2021}
\preprint{NORDITA-2021-078}

\begin{abstract} 
In Chain Inflation the universe tunnels along a series of false vacua of ever-decreasing energy.  The main goal of this paper is to embed Chain Inflation in high energy fundamental physics.  We begin by illustrating a simple effective formalism for calculating Cosmic Microwave Background (CMB) observables in Chain Inflation. Density perturbations seeding the anisotropies emerge from the probabilistic nature of tunneling (rather than from quantum fluctuations of the inflation). To obtain the correct normalization of the scalar power spectrum and the scalar spectral index, we find an upper limit on the scale of inflation at horizon crossing of CMB scales, $V_*^{1/4}< 10^{12}\:\text{GeV}$. We then provide an explicit realization of chain inflation, in which the inflaton is identified with an axion in supergravity. The axion enjoys a perturbative shift symmetry which is broken to a discrete remnant by instantons. The model, which we dub `natural chain inflation' satisfies all cosmological constraints and can be embedded into a standard $\Lambda$CDM cosmology. Our work provides a major step towards the ultraviolet completion of chain inflation in string theory.
\end{abstract}
\maketitle

\section{Introduction}

Cosmic inflation, an early accelerated epoch of the Universe, was proposed to solve several conundrums of the Hot Big Bang~\cite{Guth:1980zm}, and in addition provides density fluctuations that serve as seeds for the formation of large scale structure.
 In recent years, the paradigm most frequently studied consists of a scalar field, the inflaton, which rolls down its potential and drives the exponential expansion of space~\cite{Linde:1981mu,Albrecht:1982wi}. Quantum fluctuations of the inflaton seed the temperature anisotropies we observe in the CMB~\cite{Mukhanov:1981xt}. 

However, there exist less known, though equally successful theories of inflation,
in which the rolling of the scalar field is replaced by its tunneling from higher energy metastable vacua to lower energy minima.  During the time spent in the higher energy minimum, the potential energy drives accelerated expansion.  Guth's original ``old inflation'', with a single tunneling event, was the original tunneling model, but it suffered from a failure to reheat after inflation:
whereas pockets of the Universe do undergo the required first order phase transition, most of the Universe remains in the false vacuum and the bubbles of true vacuum never percolate or reheat.  

Several solutions to this reheating problem have been proposed, including Double Field Inflation~\cite{Adams:1990ds,Linde:1990gz} (a second rolling field gives rise to time dependence of the tunneling rate) and Chain Inflation, which is the subject of this paper.
In Chain Inflation~\cite{Freese:2004vs,Freese:2005kt,Ashoorioon:2008pj}, the Universe undergoes a series of first order phase transitions. Instead of a rolling scalar field, the inflaton tunnels along a series of metastable minima in its potential. While quantum fluctuations are suppressed due to the inflaton mass in each vacuum, the probabilistic nature of tunneling causes density perturbations in the primordial plasma which later manifest as  anisotropies in the CMB.

It was previously shown that for $\sim$10$^4$ phase transitions per e-fold of inflation, the correct amplitude of CMB temperature fluctuations is obtained~\cite{Feldstein:2006hm,Winkler:2020ape}. 
The large number of transitions automatically ensures that vacuum bubbles from individual nucleation sites quickly percolate, thus evading the ``empty universe problem''~\cite{Guth:1982pn} of old inflation. A nearly scale-invariant scalar power spectrum -- as preferred by observation -- is obtained if the Hubble rate $H$ and the tunneling rate $\Gamma$ vary (at most) slowly from vacuum to vacuum~\cite{Winkler:2020ape}.

In this work we provide simple analytic expressions to determine the CMB observables in concrete models of chain inflation. The formalism is based on our previously derived approximation of tunneling rates which replaces the thin-wall approximation for generic quasiperiodic potentials~\cite{Winkler:2020ape}.

The main goal of this paper is to illustrate an explicit supergravity realization of chain inflation in which the inflaton is identified with an axion. Two instanton terms break the axionic shift symmetry and induce a periodic potential as in Eq.~\eqref{eq:potentialmodel}. Such a setting carries profound motivation from string theory in which axions with multiple instanton terms are ubiquitous. The leading instanton term generates an overall cosine shape of the axion potential, while the subleading instanton induces a series of metastable minima by causing small wiggles in the potential. We show that the model allows for a successful regime of chain inflation in which all cosmological constraints are satisfied.

\section{CMB Observables}\label{sec:cmbobservables}

In this section we describe a quick method to derive the CMB observables of chain inflation which solely relies on the determination of the extrema in the potential. We consider a (quasi)periodic potential $V(\varphi)$ containing a series of metastable minima whose energy increases monotonically (in the regime where inflation takes place). The correct CMB normalization in chain inflation will require at least $\mathcal{O}(10^5)$ consecutive minima~\cite{Feldstein:2006hm,Winkler:2020ape}. 

\begin{figure}[b]
\begin{center}
\includegraphics[width=0.4\textwidth]{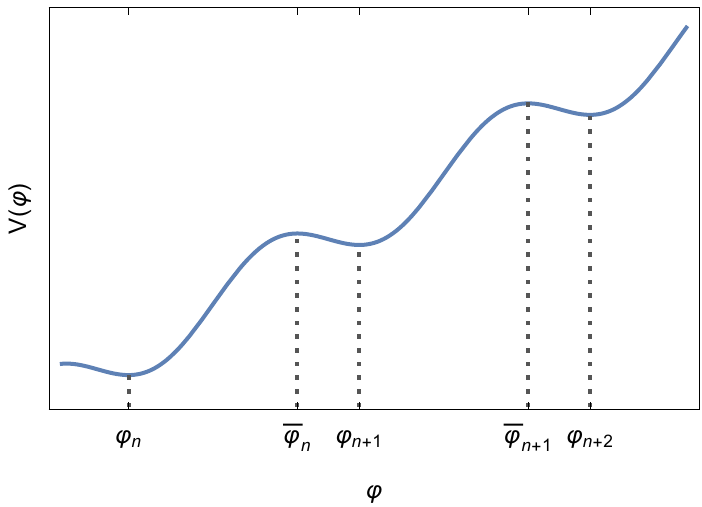}
\end{center}
\vspace{-5mm}
\caption{Typical potential in chain inflation (only a small part of the full potential is shown). Extrema are numbered by $\varphi_n$ (minima) and $\bar{\varphi}_n$ (maxima) from low to high energy.}
\label{fig:potential}
\end{figure}

First, we determine the extrema in the potential. The minima (maxima) are denoted by $\varphi_n$ ($\bar{\varphi}_n$) starting from the lowest energy minimum (maximum). The notation is illustrated in Fig.~\ref{fig:potential}. Then, the potential is expressed in the form
\begin{equation}\label{eq:potential}
V(\varphi)= \mu^3(\varphi)\: \varphi + \Lambda^4(\varphi)\cos\left(\frac{\varphi}{f(\varphi)}+\theta(\varphi)\right) +C(\varphi) \,.
\end{equation}
We assume that locally, i.e.\ between two neighboring minima, the potential can be approximated as a tilted cosine -- an assumption which is valid for a wide class of quasiperiodic potentials. The functions $\mu(\varphi)$, $\Lambda(\varphi)$, $f(\varphi)$, $\theta(\varphi)$ and $C(\varphi)$ then depend on $\varphi$, but their variation between two consecutive minima is small. 

Only $\mu$, $\Lambda$ and $f$ enter the tunneling rate between two minima as long as gravitational corrections are negligible. The value of these three functions at a specific minimum $\varphi_n$ can be determined from the following set of equations,
\begin{align}\label{eq:map}
   \mu(\varphi_n) &= \left(\frac{V(\varphi_{n+1}) - V(\varphi_{n})}{\varphi_{n+1} - \varphi_{n}}\right)^{1/3}\,,\nonumber\\
  \Lambda(\varphi_n)&=\left(\frac{V(\varphi_{n+1}) - V(\varphi_{n})}{2\pi\cos\left[ \left(  \frac{\varphi_{n+1} - \bar{\varphi}_n}{\varphi_{n+1} - \varphi_n} \right)\pi \right]} \right)^{1/4}\,,\nonumber\\
    f(\varphi_n) &=\frac{\varphi_{n+1} - \varphi_n}{2\pi}\,.
\end{align}

The tunneling rate per unit four volume $\Gamma$ for the transition between two minima reads~\cite{Coleman:1977py},
\begin{equation}\label{eq:Gamma}
  \Gamma = A \, e^{-S_E}\,,
\end{equation}
where $S_E$ is the Euclidean action of the bounce solution extrapolating between the minima. The prefactor $A$ can be derived by considering quantum fluctuations about the action of the bounce~\cite{Callan:1977pt}. For the potential~\eqref{eq:potential}, we can use the approximation we provided in~\cite{Winkler:2020ape},
\begin{equation}\label{eq:final_approximation}
\Gamma = \frac{\Lambda^8}{f^4}\,(1-x^2)\,\frac{S_E^2}{4\pi^2}\:
e^{13.15 - 15.8/x^{2.9}} \times e^{-S_E}
\end{equation}
with
\begin{align}\label{eq:bounceaction}
S_E&=\frac{f^4}{\Lambda^4}\sqrt{(1-x^2)\,(1-0.86 x^2)}\;\frac{4}{\pi} \left(\frac{12}{x}\right)^3\,,\nonumber\\
x &= \frac{f\mu^3}{\Lambda^4}\,.
\end{align}
Here and in the following we use Planck units, i.e.\ we set $M_P=1$. The above expressions allow us to determine the tunneling rate $\Gamma(\varphi_n)$ for each pair of minima $\{n+1,n\}$. The Hubble parameter at a specific minimum can be approximated as
\begin{equation}
H(\varphi_n) = \sqrt{\frac{V(\varphi_n)}{3}}\,.
\end{equation}
In the above expression we only included the potential energy of the inflaton. The rapid percolation of bubbles from individual vacuum transitions produces additional energy in the form of radiation. However, since the latter is quickly redshifted away during inflation, we neglected its contribution to the total energy budget.

In the limit of a large number of minima we can effectively treat $\Gamma$ as a continuous function of $\varphi$ by numerically interpolating $\Gamma(\varphi_n)$ between the minima. This will allow us to perform integrals and derivatives of $\Gamma$ with respect to the field in the following. Formally, we can define
\begin{equation}
\Gamma'(\varphi_n)= \frac{\Gamma(\varphi_{n+1}) - \Gamma(\varphi_{n})}{\varphi_{n+1} - \varphi_{n}}\,.
\end{equation}
The field-derivative of the Hubble parameter is obtained analogously.

We then use the result for the scalar power spectrum of chain inflation from our simulations~\cite{Winkler:2020ape},
\begin{equation}\label{eq:scalarpower}
\Delta_{\mathcal{R}}^2 \simeq 0.06 \left(\frac{\Gamma}{H^4} \right)^{-5/12}\,.
\end{equation}
The deviation of the power spectrum from scale-invariance originates from the variation of $\Gamma$ and $H$ along the inflationary trajectory. The scalar spectral index is given by~\cite{Winkler:2020ape}
\begin{equation}\label{eq:ns}
  n_s \simeq 1+ \frac{5}{12}\,\left( \frac{4\dot{H}}{H^2} - \frac{\dot{\Gamma}}{H \Gamma} \right)\,.
\end{equation}
For the quasiperiodic potentials considered here, we can express the spectral index as a function of the field value. For this purpose we use
\begin{equation}\label{eq:phidot}
\dot{\varphi}= -8.8\; \Gamma^{1/4}(\varphi) f(\varphi)\,,
\end{equation}
which follows from the expression for $\dot{\varphi}$ obtained via simulations in~\cite{Winkler:2020ape} when we additionally take into account that the field-distance between two neighboring minima is $2\pi f$. Combining~\eqref{eq:ns} and~\eqref{eq:phidot} yields,
\begin{equation}\label{eq:nsphi}
  n_s(\varphi) \simeq 1- 3.7\,f\,\Gamma^{1/4} \,\left( \frac{4H'(\varphi)}{H^2} - \frac{\Gamma'(\varphi)}{H \Gamma} \right)\,,
\end{equation}
Finally, we need to relate the field value to the number of e-folds $N$. Using $dN = H dt = (H/\dot{\varphi})\, d\varphi$, we obtain
\begin{equation}\label{eq:efolds}
N(\varphi) = \int\limits_{\varphi_0}^\varphi d\varphi' \frac{H(\varphi)}{8.8\; \Gamma^{1/4}(\varphi) f(\varphi)}\,,
\end{equation}
where $\varphi_0$ denotes the field-value at which inflation ends.

\section{Constraints on Chain Inflation}

\begin{figure*}[t]
\begin{center}
\includegraphics[width=0.845\textwidth]{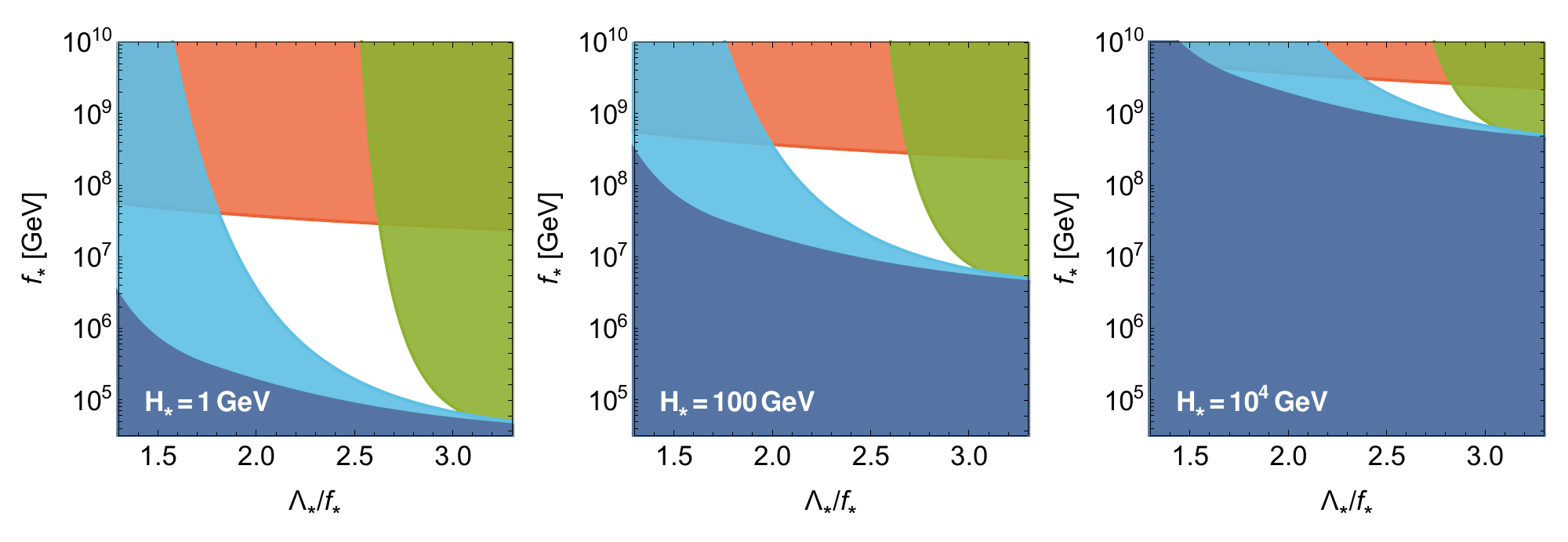}\hspace{0.007\textwidth}
\includegraphics[width=0.135\textwidth,trim={0 -1.0cm 0 0},clip]{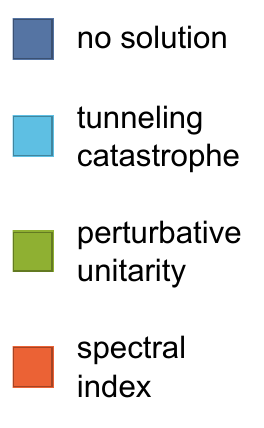}
\end{center}
\vspace{-7mm}
\caption{Constraints on chain inflation for different choices of the Hubble scale $H_*$ evaluated at horizon crossing of the CMB scales (as indicated by the star). The corresponding inflation scales are $V_*^{1/4}=\{2\times 10^9\:\text{GeV},\,2\times 10^{10}\:\text{GeV},\,2\times 10^{11}\:\text{GeV}\}$ (from left to right). At each point in the $\Lambda_*/f_*$-$f_*$-plane we fixed $\mu_*$ by requiring the correct normalization of the scalar power spectrum. Colored regions are excluded due to the constraints discussed in the text. In the white regions successful chain inflation can potentially be realized.}
\label{fig:parameters}
\end{figure*}

CMB measurements constrain the Hubble scale and the shape of the potential at the field-value $\varphi_*$. The latter is defined by the horizon crossing of the scales relevant for CMB observables, roughly $55$ e-folds before the end of inflation. We can obtain $\varphi_*$ from~\eqref{eq:efolds} by requiring $N_* \equiv N(\varphi_*) \simeq 55$. Here and in the following the star indicates that a quantity is evaluated at $\varphi_*$.

Before we turn to the model realization of chain inflation, it is useful to identify parameter combinations $\{H_*,\mu_*,\Lambda_*,f_*\}$ which can lead to successful chain inflation.

The correct amplitude of the CMB fluctuations imposes $\Delta_{\mathcal{R}}^2(\varphi_*)= 2.1\times 10^{-9}$~\cite{Planck:2018vyg} which implies (cf.~\eqref{eq:scalarpower})
\begin{equation}\label{eq:GammaH4}
\frac{\Gamma^{1/4}_*}{H_*}=3.0\times 10^4\,.
\end{equation}
This allows us to fix one of $\{H_*,\mu_*,\Lambda_*,f_*\}$ in terms of the other three parameters.

CMB data, furthermore, require a nearly scale-invariant spectrum with $n_{s*}\simeq 0.965$~\cite{Planck:2018vyg}. 
Plugging this number into~\eqref{eq:nsphi} and imposing~\eqref{eq:GammaH4} yields
\begin{equation}\label{eq:nsconstraint1}
f_*\,\left( \frac{H'_*}{H_*} - \frac{\Gamma'_*}{4 \Gamma_*} \right)= 7.9\times 10^{-8}
\end{equation}
The above constraint in its general form is not particularly useful since it not only depends on the potential parameters, but also on their derivatives. 
A more convenient condition is obtained if we additionally reject fine-tuning in the spectral index, i.e.\ exclude strong cancellations between the contributions $\propto H'_*/H_*$ and $\propto \Gamma'_*/\Gamma$. Besides unnatural, such cancellations would induce significant running of the spectral index which is experimentally excluded~\cite{Planck:2018vyg}.\footnote{The terms $\propto H'/H$ and $\propto \Gamma'/\Gamma$ have a very different dependence on $\mu$, $f$, $\Lambda$ and its derivatives. Therefore, even if a cancellation between terms $\propto H'/H$ and $\propto \Gamma'/\Gamma$ in $n_s$ occurs at $\varphi_*$ it cannot be upheld for neighboring field values and a strong running of the spectral index occurs.} We require
\begin{equation}\label{eq:notuning}
\left|\frac{H_*'}{H_*}-\frac{\Gamma'_*}{4\Gamma_*}\right| > \frac{1}{10} \left|\frac{H'_*}{H_*}\right|\,,
\end{equation}
thus tolerating at most a factor of 10 fine-tuning. Plugging~\eqref{eq:notuning} into~\eqref{eq:nsconstraint1} one obtains the condition
\begin{equation}\label{eq:nsconstraint2}
\frac{f_*\mu_*^3}{H^2_*}< 4.7\times 10^{-6}\,,
\end{equation}
where we used $H'=\mu^3/(6H)$.

In addition to the CMB constraints, the validity of our tunneling solutions requires~\cite{Cline:2011fi}
\begin{equation}\label{eq:catastrophe}
x_*<0.96\,,
\end{equation}
in order to avoid a `tunneling catastrophe', in which the inflaton overshoots the next minimum and directly tunnels to the bottom of the potential. Here $x$ is as previously defined  in Eq.(~\eqref{eq:bounceaction}).

Finally, the perturbative unitarity of the theory requires $V''''_* < 8\pi$ which translates to
\begin{equation}\label{eq:unitarity}
\frac{\Lambda^4}{f^4}\sqrt{1-x_*^2}<8\pi\,.
\end{equation}

In Fig.~\ref{fig:parameters} we applied the described constraints on the parameter space spanned by $\{H_*,\mu_*,\Lambda_*,f_*\}$. We fixed $\mu_*$ by the CMB normalization~\eqref{eq:GammaH4} and then imposed~\eqref{eq:nsconstraint2},~\eqref{eq:catastrophe} and~\eqref{eq:unitarity} on the remaining three parameters. We observe that $\Lambda_*/f_*\gtrsim 1$ is required for successful chain inflation. Furthermore, it is visible that the allowed region shrinks for increasing $H_*$. Indeed, by combining the constraints on chain inflation, we can derive an upper limit $H_*<1.5\times 10^5\:\text{GeV}$ which implies 
\begin{equation}\label{eq:maxscale}
V_*^{1/4}< 8\times 10^{11}\:\text{GeV}\,,
\end{equation}
for the scale of chain inflation. Furthermore, we recover the bound $f_*<10^{10}\:\text{GeV}$ already derived in~\cite{Winkler:2020ape}. The upper limit on $f_*$ may impose a challenge for the
realization of chain inflation in string theory, where axion decay constants often come out larger (not
too far below the Planck scale). On the other hand, several mechanisms to suppress the decay constant of string axions have been suggested (see e.g.~\cite{Svrcek:2006yi}). In the next section we will introduce an explicit supergravity model of chain inflation which can access the viable parameter space we identified.

\section{Supergravity Model}
Axions are prime candidates for the particle driving chain inflation. At the perturbative level, an axion $\varphi$ enjoys a continuous shift symmetry, i.e.\ transformations $\varphi \rightarrow \varphi + c$
leave the Lagrangian invariant. However, non-perturbative instantons break the shift-symmetry down to a discrete remnant. 
In this paper we build upon (but do not use directly) the simplest single-instanton case, in which the resulting axion potential takes the form
\begin{equation}
 V = V_0 \left[1-\cos\left(\frac{\varphi}{f}\right)\right]\,,
\end{equation}
where $f$ denotes the axion decay constant. Such a potential can give rise to natural (slow roll) inflation~\cite{Freese:1990rb}, but we cannot use it for chain inflation since it does not feature a tunneling trajectory. On the other hand, axions are generically subject to multiple instanton terms, whose interplay leads to more general periodic potentials. We will show that two instanton terms already suffice for successful chain inflation.

A particularly appealing framework for chain inflation is string theory. This is because the perturbative shift symmetry of string axions is preserved at the ultraviolet scale~\cite{Wen:1985jz,Dine:1986vd}. Hence, the axion potential is protected against quantum gravity corrections. In string theory, the non-perturbative superpotential containing matter fields $\chi_i$ can schematically be written as (see e.g.~\cite{Blumenhagen:2009qh})
\begin{equation}\label{eq:mattersuperpotential}
W\supset \prod\limits_i \chi_i\: e^{-\mathcal{S}(T_i)}\,,
\end{equation}
with $\mathcal{S}$ denoting the instanton action depending on the moduli $T_i$. If we, for example, consider world-sheet instantons~\cite{Dine:1986zy} as the microscopic origin of the non-perturbative effect, $T_i$ could be identified with K\"{a}hler (or complex structure) moduli.

We consider a simplified supergravity model of this type containing two chiral superfields $T$ and $\chi$ with the superpotential and K\"{a}hler potential~\cite{Kallosh:2014vja}
\begin{align}\label{eq:sugramodel}
 W &= \chi\left(Ae^{-aT} + Be^{-aT} - C\right)\,,\nonumber\\
 K &= k_1(\bar{\chi}\chi) + k_2(\bar{T}+T)\,,
\end{align}
where $A$, $B$, $C$ are constants and $k_{1,2}$ functions of the fields. The K\"{a}hler potential~\eqref{eq:sugramodel} only depends on the real part of $T$ and, hence, contains a shift symmetry in $\varphi\propto\text{Im}\,T$ which will become the axion driving chain inflation. A non-vanishing axion potential is generated by the two instanton terms in the superpotential.

\begin{figure*}[t]
\begin{center}
\includegraphics[width=0.43\textwidth]{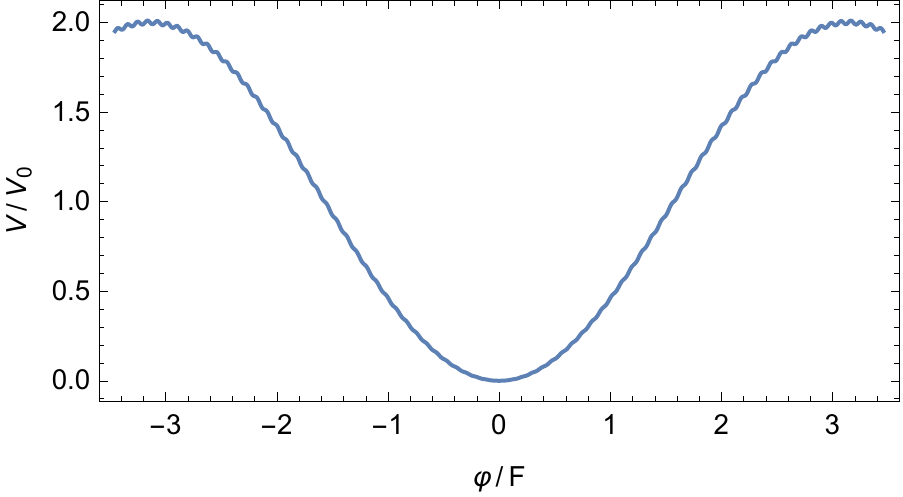}\hspace{3mm}
\includegraphics[width=0.43\textwidth]{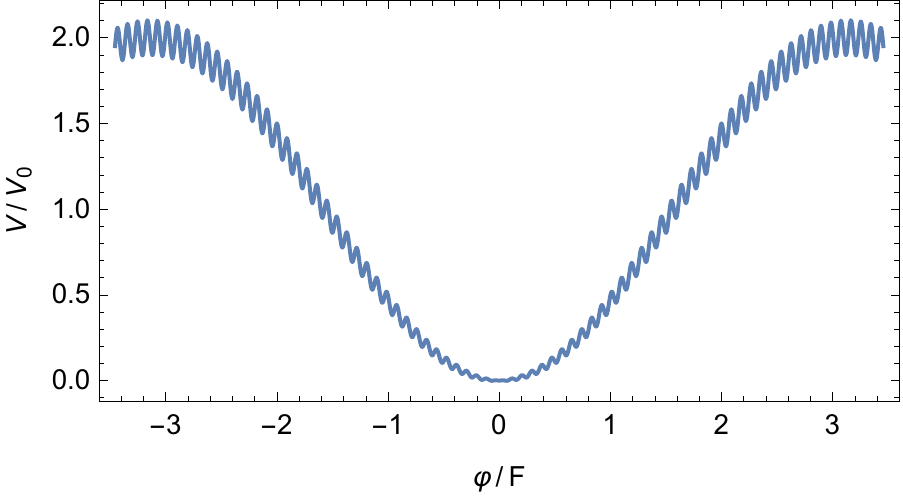}
\end{center}
\vspace{-7mm}
\caption{Illustration of the axion potential in the supergravity model. The leading instanton induces an overall cosine potential, the subleading instanton induces additional wiggles. Depending on the size of the wiggles (choice of $\delta$) the potential may give rise to slow roll inflation (left panel) or natural chain inflation (right panel).}
\label{fig:sugraaxionpotential}
\end{figure*}

The second superfield $\chi$ ensures the presence of a supersymmetric ground state with vanishing vacuum energy in which the universe ends up after inflation. The latter is located at $\chi=0$, $T=T_0$, where $T_0$ is implicitly defined by $Ae^{-aT_0} + Be^{-aT_0} =C$. We assume that $\chi$ remains fixed at $\chi=0$ during inflation either through an appropriate K\"{a}hler potential\footnote{For example, $k_1=|\chi|^2-|\chi|^4/\Lambda^2$ with $\Lambda<1$ strongly stabilizes $\chi$ at $\chi=0$~\cite{Dine:1983ys,Kallosh:2010xz}. Such a K\"{a}hler potential can be generated after integrating out Yukawa interactions of $\chi$ with heavy fields at the scale $\Lambda$~\cite{Grisaru:1996ve,Komargodski:2009rz}.} or by identifying it as a Nilpotent field~\cite{Ferrara:2014kva}. Along the inflationary trajectory, the scalar potential thus takes the form
\begin{equation}
V=e^{k_2} \,|\partial_\chi W|^2\,.
\end{equation}
We focus on the case $Ae^{-aT_0}\gg Be^{-bT_0}$ and neglect terms of $\mathcal{O}(B^2)$. Furthermore, we set $\text{Re}\,T\simeq T_0$ during inflation.\footnote{The dynamics of $\text{Re}\,T$ during inflation is controlled by its K\"{a}hler potential $k_2$. See~\cite{Kappl:2015pxa} for a string-motivated choice of $k_2$ which fixes $T\simeq T_0$ during inflation.} This assumption is not crucial for realizing chain inflation but leads to a simple analytic axion potential,
\begin{equation}\label{eq:potentialmodel}
V= V_0 \left[1-\cos\left(\frac{\varphi}{F}\right)+\delta\sin\left(\frac{\varphi}{2F}\right)\sin\left(\frac{\varphi}{f}\right)\right]\,.
\end{equation}
This is the potential we will use for the realization of chain inflation in the following. Notice that we introduced 
\begin{align}
V_0 &=2\,A\, C\,e^{-aT_0+k_2(2T_0)}\,,\;\;\;
\delta=-\frac{2B}{C}\,e^{-bT_0}\,,\nonumber\\
F&=\frac{\sqrt{2\,k_2''(2T_0)}}{a}\,,\;\;\;
f=\frac{\sqrt{8\,k_2''(2T_0)}}{a-2b}\,,
\end{align}
and canonically normalized the axion. Typically, the axion decay constant $F$ associated with the leading instanton is larger than the one ($f$) associated with the subleading instanton. We, hence, imply $F\gg f$ in the following.

At leading instanton order, we obtain the familiar cosine potential of the axion. The subleading instanton, however, induces additional periodic wiggles described by the term $\propto \delta$ in~\eqref{eq:potentialmodel}. An illustration of the potential can be found in Fig.~\ref{fig:sugraaxionpotential}.

\section{Natural Chain Inflation}

Potentials similar to~\eqref{eq:potentialmodel} have been considered in variants of natural inflation~\cite{Silverstein:2008sg,Kappl:2015esy,Hebecker:2015rya}. Subdominant wiggles on the potential have been found to be consistent with slow roll inflation and may even induce novel signatures in future CMB experiments~\cite{McAllister:2008hb,Winkler:2019hkh}. However, successful slow roll inflation is limited to the regime of $\delta\lesssim f^2/F^2$ since otherwise too strong scale-dependence of the scalar power spectrum would arise~\cite{Winkler:2019hkh}.

Intriguingly, while larger wiggles in the axion potential are unsuitable for slow roll inflation they can trigger successful chain inflation with the axion tunneling `from wiggle to wiggle' instead of rolling down the potential (see right panel of Fig.~\ref{fig:sugraaxionpotential}). Since the resulting scheme is a chain inflation version of natural inflation we dub it `natural chain inflation' in the following.

For calculating the CMB observables in natural chain inflation it is convenient to apply the formalism developed in section~\ref{sec:cmbobservables}. We thus need to determine $H$, $\mu$, $\Lambda$ and $f$ as a function of the inflaton (=axion) field value. For the potential of the supergravity model this can be done analytically and we obtain\footnote{Without loss of generality we assumed that inflation occurs in the range $[0,F\pi]$.}
\begin{align}
H(\varphi) &= \sqrt{\frac{2V_0}{3}}\sin\left(\frac{\varphi}{2 F}\right)\,,\nonumber\\
\mu(\varphi) &= \left(\frac{V_0}{F} \sin\left(\frac{\varphi}{F}\right)\right)^{1/3}\,,\nonumber\\
\Lambda(\varphi) &= \left(V_0\,\delta\,\sin\left(\frac{\varphi}{2F}\right)\right)^{1/4}\,,
\end{align}
while $f(\varphi)=f$ is a free constant.

For a given set of input parameters $\{V_0,F,f,\delta\}$, we first determine the field-value $\varphi_0$ when chain inflation ends. The latter is -- depending on the choice of $\delta$ -- either located at the bottom of the potential or at the transition point, where the potential becomes monotonic and the axion starts rolling.\footnote{In the second case, chain inflation could in principle be followed by an epoch of slow roll inflation. However, for sub-Planckian axion decay constants, the slow roll conditions are immediately violated once the potential becomes monotonic. Hence, we neglect this possibility and define the transition to the rolling regime as the end of inflation.} We find
\begin{equation}\label{eq:endinflation}
\varphi_0 \simeq \begin{cases} 0\phantom{2F \arccos\left(\frac{F\,\delta }{2f}\right)}\; \text{for}\;\delta \geq \frac{2 f}{F}\,,\\ 2F \arccos\left(\frac{F\,\delta }{2f}\right)\phantom{0}\; \text{for}\;\delta < \frac{2 f}{F}\,.
\end{cases}
\end{equation}
In the next step, we determine $\varphi_*$ from~\eqref{eq:efolds} which we then plug into~\eqref{eq:scalarpower} and~\eqref{eq:ns} to obtain $\Delta_{\mathcal{R}*}^{2}$ and $n_{s*}$.

\begin{figure}[t]
\begin{center}
\includegraphics[width=0.33\textwidth]{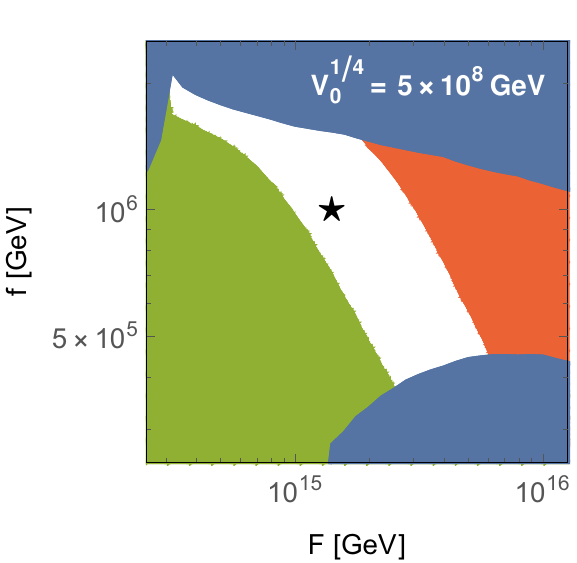}\hspace{0.5mm}
\includegraphics[width=0.13\textwidth,trim={0 -1.6cm 0 0},clip]{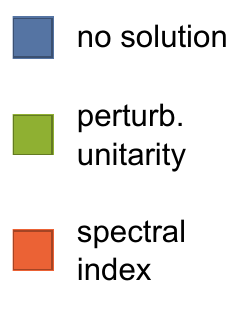}
\end{center}
\vspace{-7mm}
\caption{Constraints on natural chain inflation in terms of the axion decay constants $F$ and $f$ for fixed $V_0$. At each point in the $F$-$f$-plane we fixed $\delta$ by requiring the correct amplitude of density perturbations in the CMB. The colored regions are excluded due to the absence of a solution (blue), violation of perturbative unitarity (green) or because of a scalar spectral index which falls outside the $1\sigma$-window $n_{s*}=0.9649\pm 0.0041$ measured by Planck. In the white region all constraints are satisfied and successful chain inflation can be realized. The input parameters and CMB predictions for the benchmark point indicated by the star can be found in Tab.~\ref{tab:benchmark}.}
\label{fig:scan2}
\end{figure}

In Fig.~\ref{fig:scan2} we set $V_0^{1/4}=5\times 10^8\:\text{GeV}$ and scan over the two axion decay constants $F$ and $f$. At each parameter point $\delta$ is fixed by imposing the correct amplitude of the scalar power spectrum (cf.~\eqref{eq:GammaH4}).
Requiring the observed spectral index $n_{s*}=0.9649\pm 0.0042$~\cite{Planck:2018vyg} and perturbative unitarity then considerably narrows down the parameter space. However, the white region in the figure survives all constraints. A parameter example (indicated by the star in Fig.~\ref{fig:scan2}) and the corresponding CMB predictions can be found in Tab.~\ref{tab:benchmark}.

By scanning over $V_0$ we can then also derive an upper limit on the scale of inflation. We obtain
\begin{equation}
V_*^{1/4} < 10^{11}\:\text{GeV}\,,
\end{equation}
consistent with the general constraint for (quasi)periodic potentials in Eq.~\eqref{eq:maxscale}.

We have thus found a successful particle physics implementation of chain inflation which is fully consistent with a standard $\Lambda$CDM cosmology. Our model of natural chain inflation can be realized with a generic axion in supergravity. 

\begin{table}[t]
\begin{center}
\begin{tabular}{|c|c|c|c|}
\hline
  &&&\\[-4mm]
  $\;\;\;V_0^{1/4}$~[GeV]$\;\;\;$ & $\;\;\;\;F$~[GeV]$\;\;\;\;$ & $\;\;\;\;f$~[GeV]$\;\;\;\;$ & $\;\;\;\;\;\;\;\;\;\;\delta\;\;\;\;\;\;\;\;\;\;$\\
  &&&\\[-4mm]
  \hline
  $5\times 10^8$    & $1.4 \times 10^{15}$ & $10^6$ & $4.8\times 10^{-10}$ \\
  &&&\\[-4mm]
  \hline\hline &&& \\[-4mm]
  $\Delta_{\mathcal{R}}^2(\varphi_*)$ & $n_{s*}$ & $H_*$~[GeV] & $V_*^{1/4}$~[GeV] \\
  &&&\\[-4mm]
  \hline
  $2.1\times 10^{-9}$ & $ 0.966$ & $0.078$ & $5.8\times 10^8$\\
  \hline
\end{tabular}
\end{center}
\vspace{-0.4cm}
\caption{Input parameters and CMB predictions for the benchmark point indicated in Fig.~\ref{fig:scan2}.}
\label{tab:benchmark}
\end{table}

\section{Conclusion and Outlook}

In chain inflation the expansion of space is driven by the energy of false vacua which subsequently decay through quantum tunneling. While the emerging picture is markedly different from slow roll inflation, both theories are equally consistent with all cosmological constraints.

In this work we developed a simple formalism for calculating CMB observables in chain inflation. The formalism applies to a wide class of inflation models with (quasi)periodic potentials and solely relies on the determination of the extrema in the potential.

We then introduced a concrete particle physics model of chain inflation in the framework of supergravity. The inflaton is identified with an axion whose shift symmetry is non-perturbatively broken by two instanton terms. The resulting potential is given in Eq.~\eqref{eq:potentialmodel}. Reflecting its resemblance with natural inflation we denoted the scheme `natural chain inflation'. We showed explicitly, that the model can fit the observed amplitude and spectral index of scalar perturbations. It is, hence, fully compatible with the cosmological standard model $\Lambda$CDM.

An important subject of future research is the observational distinction of (natural) chain inflation from slow roll inflation. Promising signatures could, for example, arise in the form of gravity waves~\cite{Ashoorioon:2008nh}. The latter emerge from the collisions of bubble walls triggered by the vacuum decays in chain inflation. Further insights may also be gained by embedding natural chain inflation into a full particle physics framework of the early universe.

In particular, it will be extremely interesting to search for an explicit realization of natural chain inflation in string theory. This avenue appears fruitful since the main ingredients, axions and instantons, are ubiquitous in the string landscape. Furthermore, several key features of natural chain inflation align with general arguments on the theory of quantum gravity which were formulated in the swampland conjectures. First -- in contrast to the related slow roll models -- natural chain inflation does not require any trans-Planckian axion decay constant and, hence, satisfies the strongest version of the weak gravity conjecture~\cite{Arkani-Hamed:2006emk,Montero:2015ofa,Brown:2015iha}. Second, a low inflation scale $V_*^{1/4} < 10^{11}\:\text{GeV}$ is preferred, indicating that the trans-Planckian censorship conjecture is satisfied in a large part of the model parameter space~\cite{Bedroya:2019snp,Bedroya:2019tba}. Finally, CMB constraints restrict natural chain inflation to a regime of highly unstable de Sitter spaces which were considered as a loophole~\cite{Andriot:2018wzk,Garg:2018reu,Bedroya:2020rac} to the de Sitter conjecture~\cite{Obied:2018sgi}. In this light, there exist very exciting prospects for the ultraviolet completion of natural chain inflation.

\section*{Acknowledgments}
K.F.\ is Jeff \& Gail Kodosky Endowed Chair in Physics at the University of Texas at Austin, and K.F.\ and M.W.\ are grateful for support via this Chair.  K.F., A.L.\ and M.W.\ acknowledge support by the Swedish Research Council (Contract No. 638-2013-8993).
K.F. and M.W. are grateful for support from the U.S. Department of Energy, Office of Science, Office of High Energy Physics program under Award Number DE-SC-0002424.

\bibliography{chain}

\begin{thebibliography}{44}%
\makeatletter
\providecommand \@ifxundefined [1]{%
 \@ifx{#1\undefined}
}%
\providecommand \@ifnum [1]{%
 \ifnum #1\expandafter \@firstoftwo
 \else \expandafter \@secondoftwo
 \fi
}%
\providecommand \@ifx [1]{%
 \ifx #1\expandafter \@firstoftwo
 \else \expandafter \@secondoftwo
 \fi
}%
\providecommand \natexlab [1]{#1}%
\providecommand \enquote  [1]{``#1''}%
\providecommand \bibnamefont  [1]{#1}%
\providecommand \bibfnamefont [1]{#1}%
\providecommand \citenamefont [1]{#1}%
\providecommand \href@noop [0]{\@secondoftwo}%
\providecommand \href [0]{\begingroup \@sanitize@url \@href}%
\providecommand \@href[1]{\@@startlink{#1}\@@href}%
\providecommand \@@href[1]{\endgroup#1\@@endlink}%
\providecommand \@sanitize@url [0]{\catcode `\\12\catcode `\$12\catcode
  `\&12\catcode `\#12\catcode `\^12\catcode `\_12\catcode `\%12\relax}%
\providecommand \@@startlink[1]{}%
\providecommand \@@endlink[0]{}%
\providecommand \url  [0]{\begingroup\@sanitize@url \@url }%
\providecommand \@url [1]{\endgroup\@href {#1}{\urlprefix }}%
\providecommand \urlprefix  [0]{URL }%
\providecommand \Eprint [0]{\href }%
\providecommand \doibase [0]{http://dx.doi.org/}%
\providecommand \selectlanguage [0]{\@gobble}%
\providecommand \bibinfo  [0]{\@secondoftwo}%
\providecommand \bibfield  [0]{\@secondoftwo}%
\providecommand \translation [1]{[#1]}%
\providecommand \BibitemOpen [0]{}%
\providecommand \bibitemStop [0]{}%
\providecommand \bibitemNoStop [0]{.\EOS\space}%
\providecommand \EOS [0]{\spacefactor3000\relax}%
\providecommand \BibitemShut  [1]{\csname bibitem#1\endcsname}%
\let\auto@bib@innerbib\@empty
\bibitem [{\citenamefont {Guth}(1981)}]{Guth:1980zm}%
  \BibitemOpen
  \bibfield  {author} {\bibinfo {author} {\bibfnamefont {A.~H.}\ \bibnamefont
  {Guth}},\ }\href {\doibase 10.1103/PhysRevD.23.347} {\bibfield  {journal}
  {\bibinfo  {journal} {Phys. Rev. D}\ }\textbf {\bibinfo {volume} {23}},\
  \bibinfo {pages} {347} (\bibinfo {year} {1981})}\BibitemShut {NoStop}%
\bibitem [{\citenamefont {Linde}(1982)}]{Linde:1981mu}%
  \BibitemOpen
  \bibfield  {author} {\bibinfo {author} {\bibfnamefont {A.~D.}\ \bibnamefont
  {Linde}},\ }\href {\doibase 10.1016/0370-2693(82)91219-9} {\bibfield
  {journal} {\bibinfo  {journal} {Phys. Lett. B}\ }\textbf {\bibinfo {volume}
  {108}},\ \bibinfo {pages} {389} (\bibinfo {year} {1982})}\BibitemShut
  {NoStop}%
\bibitem [{\citenamefont {Albrecht}\ and\ \citenamefont
  {Steinhardt}(1982)}]{Albrecht:1982wi}%
  \BibitemOpen
  \bibfield  {author} {\bibinfo {author} {\bibfnamefont {A.}~\bibnamefont
  {Albrecht}}\ and\ \bibinfo {author} {\bibfnamefont {P.~J.}\ \bibnamefont
  {Steinhardt}},\ }\href {\doibase 10.1103/PhysRevLett.48.1220} {\bibfield
  {journal} {\bibinfo  {journal} {Phys. Rev. Lett.}\ }\textbf {\bibinfo
  {volume} {48}},\ \bibinfo {pages} {1220} (\bibinfo {year}
  {1982})}\BibitemShut {NoStop}%
\bibitem [{\citenamefont {Mukhanov}\ and\ \citenamefont
  {Chibisov}(1981)}]{Mukhanov:1981xt}%
  \BibitemOpen
  \bibfield  {author} {\bibinfo {author} {\bibfnamefont {V.~F.}\ \bibnamefont
  {Mukhanov}}\ and\ \bibinfo {author} {\bibfnamefont {G.~V.}\ \bibnamefont
  {Chibisov}},\ }\href@noop {} {\bibfield  {journal} {\bibinfo  {journal} {JETP
  Lett.}\ }\textbf {\bibinfo {volume} {33}},\ \bibinfo {pages} {532} (\bibinfo
  {year} {1981})}\BibitemShut {NoStop}%
\bibitem [{\citenamefont {Adams}\ and\ \citenamefont
  {Freese}(1991)}]{Adams:1990ds}%
  \BibitemOpen
  \bibfield  {author} {\bibinfo {author} {\bibfnamefont {F.~C.}\ \bibnamefont
  {Adams}}\ and\ \bibinfo {author} {\bibfnamefont {K.}~\bibnamefont {Freese}},\
  }\href {\doibase 10.1103/PhysRevD.43.353} {\bibfield  {journal} {\bibinfo
  {journal} {Phys. Rev. D}\ }\textbf {\bibinfo {volume} {43}},\ \bibinfo
  {pages} {353} (\bibinfo {year} {1991})},\ \Eprint
  {http://arxiv.org/abs/hep-ph/0504135} {arXiv:hep-ph/0504135} \BibitemShut
  {NoStop}%
\bibitem [{\citenamefont {Linde}(1990)}]{Linde:1990gz}%
  \BibitemOpen
  \bibfield  {author} {\bibinfo {author} {\bibfnamefont {A.~D.}\ \bibnamefont
  {Linde}},\ }\href {\doibase 10.1016/0370-2693(90)90521-7} {\bibfield
  {journal} {\bibinfo  {journal} {Phys. Lett. B}\ }\textbf {\bibinfo {volume}
  {249}},\ \bibinfo {pages} {18} (\bibinfo {year} {1990})}\BibitemShut
  {NoStop}%
\bibitem [{\citenamefont {Freese}\ and\ \citenamefont
  {Spolyar}(2005)}]{Freese:2004vs}%
  \BibitemOpen
  \bibfield  {author} {\bibinfo {author} {\bibfnamefont {K.}~\bibnamefont
  {Freese}}\ and\ \bibinfo {author} {\bibfnamefont {D.}~\bibnamefont
  {Spolyar}},\ }\href {\doibase 10.1088/1475-7516/2005/07/007} {\bibfield
  {journal} {\bibinfo  {journal} {JCAP}\ }\textbf {\bibinfo {volume} {07}},\
  \bibinfo {pages} {007} (\bibinfo {year} {2005})},\ \Eprint
  {http://arxiv.org/abs/hep-ph/0412145} {arXiv:hep-ph/0412145} \BibitemShut
  {NoStop}%
\bibitem [{\citenamefont {Freese}\ \emph {et~al.}(2005)\citenamefont {Freese},
  \citenamefont {Liu},\ and\ \citenamefont {Spolyar}}]{Freese:2005kt}%
  \BibitemOpen
  \bibfield  {author} {\bibinfo {author} {\bibfnamefont {K.}~\bibnamefont
  {Freese}}, \bibinfo {author} {\bibfnamefont {J.~T.}\ \bibnamefont {Liu}}, \
  and\ \bibinfo {author} {\bibfnamefont {D.}~\bibnamefont {Spolyar}},\ }\href
  {\doibase 10.1103/PhysRevD.72.123521} {\bibfield  {journal} {\bibinfo
  {journal} {Phys. Rev. D}\ }\textbf {\bibinfo {volume} {72}},\ \bibinfo
  {pages} {123521} (\bibinfo {year} {2005})},\ \Eprint
  {http://arxiv.org/abs/hep-ph/0502177} {arXiv:hep-ph/0502177} \BibitemShut
  {NoStop}%
\bibitem [{\citenamefont {Ashoorioon}\ \emph {et~al.}(2009)\citenamefont
  {Ashoorioon}, \citenamefont {Freese},\ and\ \citenamefont
  {Liu}}]{Ashoorioon:2008pj}%
  \BibitemOpen
  \bibfield  {author} {\bibinfo {author} {\bibfnamefont {A.}~\bibnamefont
  {Ashoorioon}}, \bibinfo {author} {\bibfnamefont {K.}~\bibnamefont {Freese}},
  \ and\ \bibinfo {author} {\bibfnamefont {J.~T.}\ \bibnamefont {Liu}},\ }\href
  {\doibase 10.1103/PhysRevD.79.067302} {\bibfield  {journal} {\bibinfo
  {journal} {Phys. Rev. D}\ }\textbf {\bibinfo {volume} {79}},\ \bibinfo
  {pages} {067302} (\bibinfo {year} {2009})},\ \Eprint
  {http://arxiv.org/abs/0810.0228} {arXiv:0810.0228 [hep-ph]} \BibitemShut
  {NoStop}%
\bibitem [{\citenamefont {Feldstein}\ and\ \citenamefont
  {Tweedie}(2007)}]{Feldstein:2006hm}%
  \BibitemOpen
  \bibfield  {author} {\bibinfo {author} {\bibfnamefont {B.}~\bibnamefont
  {Feldstein}}\ and\ \bibinfo {author} {\bibfnamefont {B.}~\bibnamefont
  {Tweedie}},\ }\href {\doibase 10.1088/1475-7516/2007/04/020} {\bibfield
  {journal} {\bibinfo  {journal} {JCAP}\ }\textbf {\bibinfo {volume} {04}},\
  \bibinfo {pages} {020} (\bibinfo {year} {2007})},\ \Eprint
  {http://arxiv.org/abs/hep-ph/0611286} {arXiv:hep-ph/0611286} \BibitemShut
  {NoStop}%
\bibitem [{\citenamefont {Winkler}\ and\ \citenamefont
  {Freese}(2021)}]{Winkler:2020ape}%
  \BibitemOpen
  \bibfield  {author} {\bibinfo {author} {\bibfnamefont {M.~W.}\ \bibnamefont
  {Winkler}}\ and\ \bibinfo {author} {\bibfnamefont {K.}~\bibnamefont
  {Freese}},\ }\href {\doibase 10.1103/PhysRevD.103.043511} {\bibfield
  {journal} {\bibinfo  {journal} {Phys. Rev. D}\ }\textbf {\bibinfo {volume}
  {103}},\ \bibinfo {pages} {043511} (\bibinfo {year} {2021})},\ \Eprint
  {http://arxiv.org/abs/2011.12980} {arXiv:2011.12980 [hep-th]} \BibitemShut
  {NoStop}%
\bibitem [{\citenamefont {Guth}\ and\ \citenamefont
  {Weinberg}(1983)}]{Guth:1982pn}%
  \BibitemOpen
  \bibfield  {author} {\bibinfo {author} {\bibfnamefont {A.~H.}\ \bibnamefont
  {Guth}}\ and\ \bibinfo {author} {\bibfnamefont {E.~J.}\ \bibnamefont
  {Weinberg}},\ }\href {\doibase 10.1016/0550-3213(83)90307-3} {\bibfield
  {journal} {\bibinfo  {journal} {Nucl. Phys. B}\ }\textbf {\bibinfo {volume}
  {212}},\ \bibinfo {pages} {321} (\bibinfo {year} {1983})}\BibitemShut
  {NoStop}%
\bibitem [{\citenamefont {Coleman}(1977)}]{Coleman:1977py}%
  \BibitemOpen
  \bibfield  {author} {\bibinfo {author} {\bibfnamefont {S.~R.}\ \bibnamefont
  {Coleman}},\ }\href {\doibase 10.1103/PhysRevD.16.1248} {\bibfield  {journal}
  {\bibinfo  {journal} {Phys. Rev. D}\ }\textbf {\bibinfo {volume} {15}},\
  \bibinfo {pages} {2929} (\bibinfo {year} {1977})},\ \bibinfo {note}
  {[Erratum: Phys.Rev.D 16, 1248 (1977)]}\BibitemShut {NoStop}%
\bibitem [{\citenamefont {Callan}\ and\ \citenamefont
  {Coleman}(1977)}]{Callan:1977pt}%
  \BibitemOpen
  \bibfield  {author} {\bibinfo {author} {\bibfnamefont {J.}~\bibnamefont
  {Callan}, \bibfnamefont {Curtis~G.}}\ and\ \bibinfo {author} {\bibfnamefont
  {S.~R.}\ \bibnamefont {Coleman}},\ }\href {\doibase 10.1103/PhysRevD.16.1762}
  {\bibfield  {journal} {\bibinfo  {journal} {Phys. Rev. D}\ }\textbf {\bibinfo
  {volume} {16}},\ \bibinfo {pages} {1762} (\bibinfo {year}
  {1977})}\BibitemShut {NoStop}%
\bibitem [{\citenamefont {Aghanim}\ \emph {et~al.}(2020)\citenamefont {Aghanim}
  \emph {et~al.}}]{Planck:2018vyg}%
  \BibitemOpen
  \bibfield  {author} {\bibinfo {author} {\bibfnamefont {N.}~\bibnamefont
  {Aghanim}} \emph {et~al.} (\bibinfo {collaboration} {Planck}),\ }\href
  {\doibase 10.1051/0004-6361/201833910} {\bibfield  {journal} {\bibinfo
  {journal} {Astron. Astrophys.}\ }\textbf {\bibinfo {volume} {641}},\ \bibinfo
  {pages} {A6} (\bibinfo {year} {2020})},\ \bibinfo {note} {[Erratum:
  Astron.Astrophys. 652, C4 (2021)]},\ \Eprint
  {http://arxiv.org/abs/1807.06209} {arXiv:1807.06209 [astro-ph.CO]}
  \BibitemShut {NoStop}%
\bibitem [{\citenamefont {Cline}\ \emph {et~al.}(2011)\citenamefont {Cline},
  \citenamefont {Moore},\ and\ \citenamefont {Wang}}]{Cline:2011fi}%
  \BibitemOpen
  \bibfield  {author} {\bibinfo {author} {\bibfnamefont {J.~M.}\ \bibnamefont
  {Cline}}, \bibinfo {author} {\bibfnamefont {G.~D.}\ \bibnamefont {Moore}}, \
  and\ \bibinfo {author} {\bibfnamefont {Y.}~\bibnamefont {Wang}},\ }\href
  {\doibase 10.1088/1475-7516/2011/08/032} {\bibfield  {journal} {\bibinfo
  {journal} {JCAP}\ }\textbf {\bibinfo {volume} {08}},\ \bibinfo {pages} {032}
  (\bibinfo {year} {2011})},\ \Eprint {http://arxiv.org/abs/1106.2188}
  {arXiv:1106.2188 [hep-th]} \BibitemShut {NoStop}%
\bibitem [{\citenamefont {Svrcek}\ and\ \citenamefont
  {Witten}(2006)}]{Svrcek:2006yi}%
  \BibitemOpen
  \bibfield  {author} {\bibinfo {author} {\bibfnamefont {P.}~\bibnamefont
  {Svrcek}}\ and\ \bibinfo {author} {\bibfnamefont {E.}~\bibnamefont
  {Witten}},\ }\href {\doibase 10.1088/1126-6708/2006/06/051} {\bibfield
  {journal} {\bibinfo  {journal} {JHEP}\ }\textbf {\bibinfo {volume} {06}},\
  \bibinfo {pages} {051} (\bibinfo {year} {2006})},\ \Eprint
  {http://arxiv.org/abs/hep-th/0605206} {arXiv:hep-th/0605206} \BibitemShut
  {NoStop}%
\bibitem [{\citenamefont {Freese}\ \emph {et~al.}(1990)\citenamefont {Freese},
  \citenamefont {Frieman},\ and\ \citenamefont {Olinto}}]{Freese:1990rb}%
  \BibitemOpen
  \bibfield  {author} {\bibinfo {author} {\bibfnamefont {K.}~\bibnamefont
  {Freese}}, \bibinfo {author} {\bibfnamefont {J.~A.}\ \bibnamefont {Frieman}},
  \ and\ \bibinfo {author} {\bibfnamefont {A.~V.}\ \bibnamefont {Olinto}},\
  }\href {\doibase 10.1103/PhysRevLett.65.3233} {\bibfield  {journal} {\bibinfo
   {journal} {Phys. Rev. Lett.}\ }\textbf {\bibinfo {volume} {65}},\ \bibinfo
  {pages} {3233} (\bibinfo {year} {1990})}\BibitemShut {NoStop}%
\bibitem [{\citenamefont {Wen}\ and\ \citenamefont
  {Witten}(1986)}]{Wen:1985jz}%
  \BibitemOpen
  \bibfield  {author} {\bibinfo {author} {\bibfnamefont {X.~G.}\ \bibnamefont
  {Wen}}\ and\ \bibinfo {author} {\bibfnamefont {E.}~\bibnamefont {Witten}},\
  }\href {\doibase 10.1016/0370-2693(86)91587-X} {\bibfield  {journal}
  {\bibinfo  {journal} {Phys. Lett. B}\ }\textbf {\bibinfo {volume} {166}},\
  \bibinfo {pages} {397} (\bibinfo {year} {1986})}\BibitemShut {NoStop}%
\bibitem [{\citenamefont {Dine}\ and\ \citenamefont
  {Seiberg}(1986)}]{Dine:1986vd}%
  \BibitemOpen
  \bibfield  {author} {\bibinfo {author} {\bibfnamefont {M.}~\bibnamefont
  {Dine}}\ and\ \bibinfo {author} {\bibfnamefont {N.}~\bibnamefont {Seiberg}},\
  }\href {\doibase 10.1103/PhysRevLett.57.2625} {\bibfield  {journal} {\bibinfo
   {journal} {Phys. Rev. Lett.}\ }\textbf {\bibinfo {volume} {57}},\ \bibinfo
  {pages} {2625} (\bibinfo {year} {1986})}\BibitemShut {NoStop}%
\bibitem [{\citenamefont {Blumenhagen}\ \emph {et~al.}(2009)\citenamefont
  {Blumenhagen}, \citenamefont {Cvetic}, \citenamefont {Kachru},\ and\
  \citenamefont {Weigand}}]{Blumenhagen:2009qh}%
  \BibitemOpen
  \bibfield  {author} {\bibinfo {author} {\bibfnamefont {R.}~\bibnamefont
  {Blumenhagen}}, \bibinfo {author} {\bibfnamefont {M.}~\bibnamefont {Cvetic}},
  \bibinfo {author} {\bibfnamefont {S.}~\bibnamefont {Kachru}}, \ and\ \bibinfo
  {author} {\bibfnamefont {T.}~\bibnamefont {Weigand}},\ }\href {\doibase
  10.1146/annurev.nucl.010909.083113} {\bibfield  {journal} {\bibinfo
  {journal} {Ann. Rev. Nucl. Part. Sci.}\ }\textbf {\bibinfo {volume} {59}},\
  \bibinfo {pages} {269} (\bibinfo {year} {2009})},\ \Eprint
  {http://arxiv.org/abs/0902.3251} {arXiv:0902.3251 [hep-th]} \BibitemShut
  {NoStop}%
\bibitem [{\citenamefont {Dine}\ \emph {et~al.}(1986)\citenamefont {Dine},
  \citenamefont {Seiberg}, \citenamefont {Wen},\ and\ \citenamefont
  {Witten}}]{Dine:1986zy}%
  \BibitemOpen
  \bibfield  {author} {\bibinfo {author} {\bibfnamefont {M.}~\bibnamefont
  {Dine}}, \bibinfo {author} {\bibfnamefont {N.}~\bibnamefont {Seiberg}},
  \bibinfo {author} {\bibfnamefont {X.~G.}\ \bibnamefont {Wen}}, \ and\
  \bibinfo {author} {\bibfnamefont {E.}~\bibnamefont {Witten}},\ }\href
  {\doibase 10.1016/0550-3213(86)90418-9} {\bibfield  {journal} {\bibinfo
  {journal} {Nucl. Phys. B}\ }\textbf {\bibinfo {volume} {278}},\ \bibinfo
  {pages} {769} (\bibinfo {year} {1986})}\BibitemShut {NoStop}%
\bibitem [{\citenamefont {Kallosh}\ \emph {et~al.}(2014)\citenamefont
  {Kallosh}, \citenamefont {Linde},\ and\ \citenamefont
  {Vercnocke}}]{Kallosh:2014vja}%
  \BibitemOpen
  \bibfield  {author} {\bibinfo {author} {\bibfnamefont {R.}~\bibnamefont
  {Kallosh}}, \bibinfo {author} {\bibfnamefont {A.}~\bibnamefont {Linde}}, \
  and\ \bibinfo {author} {\bibfnamefont {B.}~\bibnamefont {Vercnocke}},\ }\href
  {\doibase 10.1103/PhysRevD.90.041303} {\bibfield  {journal} {\bibinfo
  {journal} {Phys. Rev. D}\ }\textbf {\bibinfo {volume} {90}},\ \bibinfo
  {pages} {041303} (\bibinfo {year} {2014})},\ \Eprint
  {http://arxiv.org/abs/1404.6244} {arXiv:1404.6244 [hep-th]} \BibitemShut
  {NoStop}%
\bibitem [{\citenamefont {Dine}\ \emph {et~al.}(1984)\citenamefont {Dine},
  \citenamefont {Fischler},\ and\ \citenamefont {Nemeschansky}}]{Dine:1983ys}%
  \BibitemOpen
  \bibfield  {author} {\bibinfo {author} {\bibfnamefont {M.}~\bibnamefont
  {Dine}}, \bibinfo {author} {\bibfnamefont {W.}~\bibnamefont {Fischler}}, \
  and\ \bibinfo {author} {\bibfnamefont {D.}~\bibnamefont {Nemeschansky}},\
  }\href {\doibase 10.1016/0370-2693(84)91174-2} {\bibfield  {journal}
  {\bibinfo  {journal} {Phys. Lett. B}\ }\textbf {\bibinfo {volume} {136}},\
  \bibinfo {pages} {169} (\bibinfo {year} {1984})}\BibitemShut {NoStop}%
\bibitem [{\citenamefont {Kallosh}\ \emph {et~al.}(2011)\citenamefont
  {Kallosh}, \citenamefont {Linde},\ and\ \citenamefont
  {Rube}}]{Kallosh:2010xz}%
  \BibitemOpen
  \bibfield  {author} {\bibinfo {author} {\bibfnamefont {R.}~\bibnamefont
  {Kallosh}}, \bibinfo {author} {\bibfnamefont {A.}~\bibnamefont {Linde}}, \
  and\ \bibinfo {author} {\bibfnamefont {T.}~\bibnamefont {Rube}},\ }\href
  {\doibase 10.1103/PhysRevD.83.043507} {\bibfield  {journal} {\bibinfo
  {journal} {Phys. Rev. D}\ }\textbf {\bibinfo {volume} {83}},\ \bibinfo
  {pages} {043507} (\bibinfo {year} {2011})},\ \Eprint
  {http://arxiv.org/abs/1011.5945} {arXiv:1011.5945 [hep-th]} \BibitemShut
  {NoStop}%
\bibitem [{\citenamefont {Grisaru}\ \emph {et~al.}(1996)\citenamefont
  {Grisaru}, \citenamefont {Rocek},\ and\ \citenamefont {von
  Unge}}]{Grisaru:1996ve}%
  \BibitemOpen
  \bibfield  {author} {\bibinfo {author} {\bibfnamefont {M.~T.}\ \bibnamefont
  {Grisaru}}, \bibinfo {author} {\bibfnamefont {M.}~\bibnamefont {Rocek}}, \
  and\ \bibinfo {author} {\bibfnamefont {R.}~\bibnamefont {von Unge}},\ }\href
  {\doibase 10.1016/0370-2693(96)00777-0} {\bibfield  {journal} {\bibinfo
  {journal} {Phys. Lett. B}\ }\textbf {\bibinfo {volume} {383}},\ \bibinfo
  {pages} {415} (\bibinfo {year} {1996})},\ \Eprint
  {http://arxiv.org/abs/hep-th/9605149} {arXiv:hep-th/9605149} \BibitemShut
  {NoStop}%
\bibitem [{\citenamefont {Komargodski}\ and\ \citenamefont
  {Seiberg}(2009)}]{Komargodski:2009rz}%
  \BibitemOpen
  \bibfield  {author} {\bibinfo {author} {\bibfnamefont {Z.}~\bibnamefont
  {Komargodski}}\ and\ \bibinfo {author} {\bibfnamefont {N.}~\bibnamefont
  {Seiberg}},\ }\href {\doibase 10.1088/1126-6708/2009/09/066} {\bibfield
  {journal} {\bibinfo  {journal} {JHEP}\ }\textbf {\bibinfo {volume} {09}},\
  \bibinfo {pages} {066} (\bibinfo {year} {2009})},\ \Eprint
  {http://arxiv.org/abs/0907.2441} {arXiv:0907.2441 [hep-th]} \BibitemShut
  {NoStop}%
\bibitem [{\citenamefont {Ferrara}\ \emph {et~al.}(2014)\citenamefont
  {Ferrara}, \citenamefont {Kallosh},\ and\ \citenamefont
  {Linde}}]{Ferrara:2014kva}%
  \BibitemOpen
  \bibfield  {author} {\bibinfo {author} {\bibfnamefont {S.}~\bibnamefont
  {Ferrara}}, \bibinfo {author} {\bibfnamefont {R.}~\bibnamefont {Kallosh}}, \
  and\ \bibinfo {author} {\bibfnamefont {A.}~\bibnamefont {Linde}},\ }\href
  {\doibase 10.1007/JHEP10(2014)143} {\bibfield  {journal} {\bibinfo  {journal}
  {JHEP}\ }\textbf {\bibinfo {volume} {10}},\ \bibinfo {pages} {143} (\bibinfo
  {year} {2014})},\ \Eprint {http://arxiv.org/abs/1408.4096} {arXiv:1408.4096
  [hep-th]} \BibitemShut {NoStop}%
\bibitem [{\citenamefont {Kappl}\ \emph {et~al.}(2015)\citenamefont {Kappl},
  \citenamefont {Nilles},\ and\ \citenamefont {Winkler}}]{Kappl:2015pxa}%
  \BibitemOpen
  \bibfield  {author} {\bibinfo {author} {\bibfnamefont {R.}~\bibnamefont
  {Kappl}}, \bibinfo {author} {\bibfnamefont {H.~P.}\ \bibnamefont {Nilles}}, \
  and\ \bibinfo {author} {\bibfnamefont {M.~W.}\ \bibnamefont {Winkler}},\
  }\href {\doibase 10.1016/j.physletb.2015.04.035} {\bibfield  {journal}
  {\bibinfo  {journal} {Phys. Lett. B}\ }\textbf {\bibinfo {volume} {746}},\
  \bibinfo {pages} {15} (\bibinfo {year} {2015})},\ \Eprint
  {http://arxiv.org/abs/1503.01777} {arXiv:1503.01777 [hep-th]} \BibitemShut
  {NoStop}%
\bibitem [{\citenamefont {Silverstein}\ and\ \citenamefont
  {Westphal}(2008)}]{Silverstein:2008sg}%
  \BibitemOpen
  \bibfield  {author} {\bibinfo {author} {\bibfnamefont {E.}~\bibnamefont
  {Silverstein}}\ and\ \bibinfo {author} {\bibfnamefont {A.}~\bibnamefont
  {Westphal}},\ }\href {\doibase 10.1103/PhysRevD.78.106003} {\bibfield
  {journal} {\bibinfo  {journal} {Phys. Rev. D}\ }\textbf {\bibinfo {volume}
  {78}},\ \bibinfo {pages} {106003} (\bibinfo {year} {2008})},\ \Eprint
  {http://arxiv.org/abs/0803.3085} {arXiv:0803.3085 [hep-th]} \BibitemShut
  {NoStop}%
\bibitem [{\citenamefont {Kappl}\ \emph {et~al.}(2016)\citenamefont {Kappl},
  \citenamefont {Nilles},\ and\ \citenamefont {Winkler}}]{Kappl:2015esy}%
  \BibitemOpen
  \bibfield  {author} {\bibinfo {author} {\bibfnamefont {R.}~\bibnamefont
  {Kappl}}, \bibinfo {author} {\bibfnamefont {H.~P.}\ \bibnamefont {Nilles}}, \
  and\ \bibinfo {author} {\bibfnamefont {M.~W.}\ \bibnamefont {Winkler}},\
  }\href {\doibase 10.1016/j.physletb.2015.12.073} {\bibfield  {journal}
  {\bibinfo  {journal} {Phys. Lett. B}\ }\textbf {\bibinfo {volume} {753}},\
  \bibinfo {pages} {653} (\bibinfo {year} {2016})},\ \Eprint
  {http://arxiv.org/abs/1511.05560} {arXiv:1511.05560 [hep-th]} \BibitemShut
  {NoStop}%
\bibitem [{\citenamefont {Hebecker}\ \emph {et~al.}(2015)\citenamefont
  {Hebecker}, \citenamefont {Mangat}, \citenamefont {Rompineve},\ and\
  \citenamefont {Witkowski}}]{Hebecker:2015rya}%
  \BibitemOpen
  \bibfield  {author} {\bibinfo {author} {\bibfnamefont {A.}~\bibnamefont
  {Hebecker}}, \bibinfo {author} {\bibfnamefont {P.}~\bibnamefont {Mangat}},
  \bibinfo {author} {\bibfnamefont {F.}~\bibnamefont {Rompineve}}, \ and\
  \bibinfo {author} {\bibfnamefont {L.~T.}\ \bibnamefont {Witkowski}},\ }\href
  {\doibase 10.1016/j.physletb.2015.07.026} {\bibfield  {journal} {\bibinfo
  {journal} {Phys. Lett. B}\ }\textbf {\bibinfo {volume} {748}},\ \bibinfo
  {pages} {455} (\bibinfo {year} {2015})},\ \Eprint
  {http://arxiv.org/abs/1503.07912} {arXiv:1503.07912 [hep-th]} \BibitemShut
  {NoStop}%
\bibitem [{\citenamefont {McAllister}\ \emph {et~al.}(2010)\citenamefont
  {McAllister}, \citenamefont {Silverstein},\ and\ \citenamefont
  {Westphal}}]{McAllister:2008hb}%
  \BibitemOpen
  \bibfield  {author} {\bibinfo {author} {\bibfnamefont {L.}~\bibnamefont
  {McAllister}}, \bibinfo {author} {\bibfnamefont {E.}~\bibnamefont
  {Silverstein}}, \ and\ \bibinfo {author} {\bibfnamefont {A.}~\bibnamefont
  {Westphal}},\ }\href {\doibase 10.1103/PhysRevD.82.046003} {\bibfield
  {journal} {\bibinfo  {journal} {Phys. Rev. D}\ }\textbf {\bibinfo {volume}
  {82}},\ \bibinfo {pages} {046003} (\bibinfo {year} {2010})},\ \Eprint
  {http://arxiv.org/abs/0808.0706} {arXiv:0808.0706 [hep-th]} \BibitemShut
  {NoStop}%
\bibitem [{\citenamefont {Winkler}\ \emph {et~al.}(2020)\citenamefont
  {Winkler}, \citenamefont {Gerbino},\ and\ \citenamefont
  {Benetti}}]{Winkler:2019hkh}%
  \BibitemOpen
  \bibfield  {author} {\bibinfo {author} {\bibfnamefont {M.~W.}\ \bibnamefont
  {Winkler}}, \bibinfo {author} {\bibfnamefont {M.}~\bibnamefont {Gerbino}}, \
  and\ \bibinfo {author} {\bibfnamefont {M.}~\bibnamefont {Benetti}},\ }\href
  {\doibase 10.1103/PhysRevD.101.083525} {\bibfield  {journal} {\bibinfo
  {journal} {Phys. Rev. D}\ }\textbf {\bibinfo {volume} {101}},\ \bibinfo
  {pages} {083525} (\bibinfo {year} {2020})},\ \Eprint
  {http://arxiv.org/abs/1911.11148} {arXiv:1911.11148 [astro-ph.CO]}
  \BibitemShut {NoStop}%
\bibitem [{\citenamefont {Ashoorioon}\ and\ \citenamefont
  {Freese}(2008)}]{Ashoorioon:2008nh}%
  \BibitemOpen
  \bibfield  {author} {\bibinfo {author} {\bibfnamefont {A.}~\bibnamefont
  {Ashoorioon}}\ and\ \bibinfo {author} {\bibfnamefont {K.}~\bibnamefont
  {Freese}},\ }\href@noop {} {\  (\bibinfo {year} {2008})},\ \Eprint
  {http://arxiv.org/abs/0811.2401} {arXiv:0811.2401 [hep-th]} \BibitemShut
  {NoStop}%
\bibitem [{\citenamefont {Arkani-Hamed}\ \emph {et~al.}(2007)\citenamefont
  {Arkani-Hamed}, \citenamefont {Motl}, \citenamefont {Nicolis},\ and\
  \citenamefont {Vafa}}]{Arkani-Hamed:2006emk}%
  \BibitemOpen
  \bibfield  {author} {\bibinfo {author} {\bibfnamefont {N.}~\bibnamefont
  {Arkani-Hamed}}, \bibinfo {author} {\bibfnamefont {L.}~\bibnamefont {Motl}},
  \bibinfo {author} {\bibfnamefont {A.}~\bibnamefont {Nicolis}}, \ and\
  \bibinfo {author} {\bibfnamefont {C.}~\bibnamefont {Vafa}},\ }\href {\doibase
  10.1088/1126-6708/2007/06/060} {\bibfield  {journal} {\bibinfo  {journal}
  {JHEP}\ }\textbf {\bibinfo {volume} {06}},\ \bibinfo {pages} {060} (\bibinfo
  {year} {2007})},\ \Eprint {http://arxiv.org/abs/hep-th/0601001}
  {arXiv:hep-th/0601001} \BibitemShut {NoStop}%
\bibitem [{\citenamefont {Montero}\ \emph {et~al.}(2015)\citenamefont
  {Montero}, \citenamefont {Uranga},\ and\ \citenamefont
  {Valenzuela}}]{Montero:2015ofa}%
  \BibitemOpen
  \bibfield  {author} {\bibinfo {author} {\bibfnamefont {M.}~\bibnamefont
  {Montero}}, \bibinfo {author} {\bibfnamefont {A.~M.}\ \bibnamefont {Uranga}},
  \ and\ \bibinfo {author} {\bibfnamefont {I.}~\bibnamefont {Valenzuela}},\
  }\href {\doibase 10.1007/JHEP08(2015)032} {\bibfield  {journal} {\bibinfo
  {journal} {JHEP}\ }\textbf {\bibinfo {volume} {08}},\ \bibinfo {pages} {032}
  (\bibinfo {year} {2015})},\ \Eprint {http://arxiv.org/abs/1503.03886}
  {arXiv:1503.03886 [hep-th]} \BibitemShut {NoStop}%
\bibitem [{\citenamefont {Brown}\ \emph {et~al.}(2015)\citenamefont {Brown},
  \citenamefont {Cottrell}, \citenamefont {Shiu},\ and\ \citenamefont
  {Soler}}]{Brown:2015iha}%
  \BibitemOpen
  \bibfield  {author} {\bibinfo {author} {\bibfnamefont {J.}~\bibnamefont
  {Brown}}, \bibinfo {author} {\bibfnamefont {W.}~\bibnamefont {Cottrell}},
  \bibinfo {author} {\bibfnamefont {G.}~\bibnamefont {Shiu}}, \ and\ \bibinfo
  {author} {\bibfnamefont {P.}~\bibnamefont {Soler}},\ }\href {\doibase
  10.1007/JHEP10(2015)023} {\bibfield  {journal} {\bibinfo  {journal} {JHEP}\
  }\textbf {\bibinfo {volume} {10}},\ \bibinfo {pages} {023} (\bibinfo {year}
  {2015})},\ \Eprint {http://arxiv.org/abs/1503.04783} {arXiv:1503.04783
  [hep-th]} \BibitemShut {NoStop}%
\bibitem [{\citenamefont {Bedroya}\ and\ \citenamefont
  {Vafa}(2020)}]{Bedroya:2019snp}%
  \BibitemOpen
  \bibfield  {author} {\bibinfo {author} {\bibfnamefont {A.}~\bibnamefont
  {Bedroya}}\ and\ \bibinfo {author} {\bibfnamefont {C.}~\bibnamefont {Vafa}},\
  }\href {\doibase 10.1007/JHEP09(2020)123} {\bibfield  {journal} {\bibinfo
  {journal} {JHEP}\ }\textbf {\bibinfo {volume} {09}},\ \bibinfo {pages} {123}
  (\bibinfo {year} {2020})},\ \Eprint {http://arxiv.org/abs/1909.11063}
  {arXiv:1909.11063 [hep-th]} \BibitemShut {NoStop}%
\bibitem [{\citenamefont {Bedroya}\ \emph
  {et~al.}(2020{\natexlab{a}})\citenamefont {Bedroya}, \citenamefont
  {Brandenberger}, \citenamefont {Loverde},\ and\ \citenamefont
  {Vafa}}]{Bedroya:2019tba}%
  \BibitemOpen
  \bibfield  {author} {\bibinfo {author} {\bibfnamefont {A.}~\bibnamefont
  {Bedroya}}, \bibinfo {author} {\bibfnamefont {R.}~\bibnamefont
  {Brandenberger}}, \bibinfo {author} {\bibfnamefont {M.}~\bibnamefont
  {Loverde}}, \ and\ \bibinfo {author} {\bibfnamefont {C.}~\bibnamefont
  {Vafa}},\ }\href {\doibase 10.1103/PhysRevD.101.103502} {\bibfield  {journal}
  {\bibinfo  {journal} {Phys. Rev. D}\ }\textbf {\bibinfo {volume} {101}},\
  \bibinfo {pages} {103502} (\bibinfo {year} {2020}{\natexlab{a}})},\ \Eprint
  {http://arxiv.org/abs/1909.11106} {arXiv:1909.11106 [hep-th]} \BibitemShut
  {NoStop}%
\bibitem [{\citenamefont {Andriot}(2018)}]{Andriot:2018wzk}%
  \BibitemOpen
  \bibfield  {author} {\bibinfo {author} {\bibfnamefont {D.}~\bibnamefont
  {Andriot}},\ }\href {\doibase 10.1016/j.physletb.2018.09.022} {\bibfield
  {journal} {\bibinfo  {journal} {Phys. Lett. B}\ }\textbf {\bibinfo {volume}
  {785}},\ \bibinfo {pages} {570} (\bibinfo {year} {2018})},\ \Eprint
  {http://arxiv.org/abs/1806.10999} {arXiv:1806.10999 [hep-th]} \BibitemShut
  {NoStop}%
\bibitem [{\citenamefont {Garg}\ and\ \citenamefont
  {Krishnan}(2019)}]{Garg:2018reu}%
  \BibitemOpen
  \bibfield  {author} {\bibinfo {author} {\bibfnamefont {S.~K.}\ \bibnamefont
  {Garg}}\ and\ \bibinfo {author} {\bibfnamefont {C.}~\bibnamefont
  {Krishnan}},\ }\href {\doibase 10.1007/JHEP11(2019)075} {\bibfield  {journal}
  {\bibinfo  {journal} {JHEP}\ }\textbf {\bibinfo {volume} {11}},\ \bibinfo
  {pages} {075} (\bibinfo {year} {2019})},\ \Eprint
  {http://arxiv.org/abs/1807.05193} {arXiv:1807.05193 [hep-th]} \BibitemShut
  {NoStop}%
\bibitem [{\citenamefont {Bedroya}\ \emph
  {et~al.}(2020{\natexlab{b}})\citenamefont {Bedroya}, \citenamefont {Montero},
  \citenamefont {Vafa},\ and\ \citenamefont {Valenzuela}}]{Bedroya:2020rac}%
  \BibitemOpen
  \bibfield  {author} {\bibinfo {author} {\bibfnamefont {A.}~\bibnamefont
  {Bedroya}}, \bibinfo {author} {\bibfnamefont {M.}~\bibnamefont {Montero}},
  \bibinfo {author} {\bibfnamefont {C.}~\bibnamefont {Vafa}}, \ and\ \bibinfo
  {author} {\bibfnamefont {I.}~\bibnamefont {Valenzuela}},\ }\href@noop {} {\
  (\bibinfo {year} {2020}{\natexlab{b}})},\ \Eprint
  {http://arxiv.org/abs/2008.07555} {arXiv:2008.07555 [hep-th]} \BibitemShut
  {NoStop}%
\bibitem [{\citenamefont {Obied}\ \emph {et~al.}(2018)\citenamefont {Obied},
  \citenamefont {Ooguri}, \citenamefont {Spodyneiko},\ and\ \citenamefont
  {Vafa}}]{Obied:2018sgi}%
  \BibitemOpen
  \bibfield  {author} {\bibinfo {author} {\bibfnamefont {G.}~\bibnamefont
  {Obied}}, \bibinfo {author} {\bibfnamefont {H.}~\bibnamefont {Ooguri}},
  \bibinfo {author} {\bibfnamefont {L.}~\bibnamefont {Spodyneiko}}, \ and\
  \bibinfo {author} {\bibfnamefont {C.}~\bibnamefont {Vafa}},\ }\href@noop {}
  {\  (\bibinfo {year} {2018})},\ \Eprint {http://arxiv.org/abs/1806.08362}
  {arXiv:1806.08362 [hep-th]} \BibitemShut {NoStop}%
\end{thebibliography}%
\end{document}